\documentclass[prl,aps,twocolumn,groupedaddress,showpacs]{revtex4}

\usepackage{bm}
\usepackage{graphicx}
\usepackage{color}

\begin{document}

\title{Band splitting and Weyl nodes in trigonal tellurium studied by angle-resolved photoemission spectroscopy and density functional theory}

\author{K. Nakayama,$^1$ M. Kuno,$^1$ K. Yamauchi,$^2$ S. Souma,$^{3,4}$\\
K. Sugawara,$^{3,4}$ T. Oguchi,$^2$ T. Sato,$^{1,4}$ and T. Takahashi$^{1,3,4}$}

\affiliation{$^1$Department of Physics, Tohoku University, Sendai 980-8578, Japan\\
$^2$Institute of Scientific and Industrial Research, Osaka University, Ibaraki, Osaka 567-0047, Japan\\
$^3$WPI Research Center, Advanced Institute for Materials Research, Tohoku University, Sendai 980-8577, Japan\\
$^4$Center for Spintronics Research Network, Tohoku University, Sendai 980-8577, Japan
}

\date{\today}

\begin{abstract}
We have performed high-resolution angle-resolved photoemission spectroscopy (ARPES) on trigonal tellurium consisting of helical chains in the crystal. Through the band-structure mapping in the three-dimensional Brillouin zone, we found a definitive evidence for the band splitting originating from the chiral nature of crystal. A direct comparison of the band dispersion between the ARPES results and the first-principles band-structure calculations suggests the presence of Weyl nodes and tiny spin-polarized hole pockets around the H point. The present result opens a pathway toward studying the interplay among crystal symmetry, band structure, and exotic physical properties in chiral crystals.
\end{abstract}


\maketitle
Chirality, the asymmetry of an object upon its mirroring, is ubiquitous in nature. In the standard model of particle physics, quantum vacuum state is regarded to be chiral and the characteristic of elementary particles (quarks and leptons) is essentially different between left-handed and right-handed counterparts. Chirality also emerges as chiral molecules, ions, and crystals, playing an important role in a wide variety of subjects such as chemistry, biology, and solid-state physics \cite{WagniereWiley2007, BarronSSR2008}. Chiral crystals are recently attracting a great deal of attention, since they are known to be useful for advanced applications such as polarization optics \cite{BarronCUP2009, TangPRL2010, KimNatNano2016}, multiferroics \cite{CheongNM2007, KimuraRMR2007}, and spintronics \cite{KoehlAPL2009, NagaosaRMP2010, GohlerScience2011}. Also, they are predicted to host peculiar optical and transport properties, essentially owing to the spin splitting of electronic energy bands originating from the space-inversion-symmetry (SIS) breaking of the crystal \cite{VorobevJETPLett1979, ShalyginPSS2012, HirayamaPRL2015, ShalyginPRB2016, YodaSciRep2016}. The chiral crystals belong to the non-symmorphic space group, and are characterized by the screw symmetry (translational plus rotational symmetry), which is currently discussed to be one of key ingredients for the realization of novel topological semimetals \cite{AgapitoPRL2013, HirayamaPRL2015, ChenPRB2016}.

   Amongst various chiral crystals, trigonal tellurium (Te) is recently attracting particular attention due to its simple crystal structure suitable for the band engineering \cite{Optical1959, Optical1967, Calc1975, ParthasarathyPRB1988, AgapitoPRL2013, PengPRB2014, HirayamaPRL2015, PengAPL2015, LinNC2016, OkuyamaASS2017}. As shown in Figs. 1(a) and 1(b), trigonal Te consists of three atoms in a unit cell with infinite helical chains arranged in a hexagonal array, which spiral around axes parallel to {\it c} ([001] axis). Depending on the right-handed or left-handed screw axis, the space group is {\it P}3$_1$21 or {\it P}3$_2$21 ({\it D}$^4_3$ or {\it D}$^6_3$) [note that the right-handed case is displayed in Figs. 1(a) and 1(b)], in which the SIS is broken. Trigonal Te is a semiconductor with a band gap of 0.32 eV \cite{Optical1959, Optical1967} in which the valence-band (VB) maxima and conduction-band (CB) minima are located at around the H point of the bulk Brillouin zone (BZ) \cite{Calc1975, AgapitoPRL2013, PengPRB2014, HirayamaPRL2015}, and it shows intriguing physical properties such as high efficiency thermoelectricity \cite{LinNC2016}, circular photon drag effect \cite{ShalyginPRB2016}, and current-induced spin polarization \cite{VorobevJETPLett1979, ShalyginPSS2012}. Under pressure, it undergoes complex structural changes and semiconductor to metal transition \cite{ParthasarathyPRB1988}. {\it Ab initio} calculations predicted a semiconductor to a strong topological insulator transition under shear strain due to the band inversion \cite{AgapitoPRL2013}. Subsequent first-principles band calculations suggested the existence of several pairs of Weyl nodes in the spin-split bulk VB and CB at ambient pressure, and a semiconductor to Weyl semimetal (WSM) transition under hydrostatic pressure \cite{HirayamaPRL2015}. While these theoretical predictions and the origin of unusual physical properties should be experimentally examined, there exist no outputs on the experimental band structure of trigonal Te. It is thus urgently required to establish its fundamental electronic states.

In this article, we report the first angle-resolved photoemission spectroscopy (ARPES) study of trigonal Te single crystal. By utilizing energy-tunable photons from synchrotron radiation, we established the VB structure in the entire bulk BZ. We found intriguing spectral features, such as the energy splitting of bulk bands at several locations in the BZ, and the possible emergence of Weyl nodes around the H point. We compare present observations with our first-principles band calculations, and discuss the consequence of our observation in relation to exotic transport and topological properties.

\begin{figure}
\includegraphics[width=3.4in]{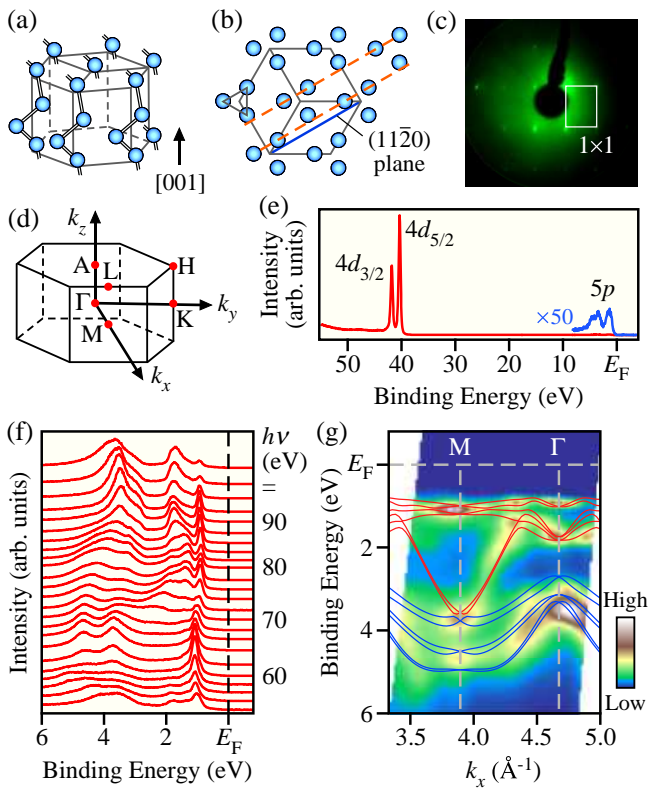}
\vspace{-0.5cm}
\caption{(Color online) (a), (b) Three-dimensional and top views of trigonal Te, respectively, consisting of right-handed helical chains. Dashed line in (b) represents mesomeric chain \cite{AgapitoPRL2013}. Cleaving plane is parallel to this chain. (c) LEED pattern of a cleaved surface measured with a primary electron energy of 80 eV. (d) Bulk BZ of trigonal Te. (e) EDC in a wide energy region measured with $h\nu$ = 80 eV. (f) Photon-energy dependence of normal-emission EDCs in the VB region measured at {\it T} = 40 K. (g) ARPES intensity in the VB region along the $\Gamma$M line plotted as a function of wave vector ($k_x$) and binding energy, together with the calculated band dispersions (red and blue curves for lone-pair and bonding 5$p$ bands, respectively).}
\end{figure}

High-quality single crystals of trigonal Te were synthesized with vapor transport method using high-purity tellurium shot (99.9999\%). ARPES measurements were performed with Scienta-Omicron SES2002 electron analyzer with energy-tunable synchrotron lights at BL28A in Photon Factory. We used photons of 50-100 eV with circular polarization. The energy and angular resolutions were set at 15-30 meV and 0.2$^\circ$, respectively. Samples were cleaved {\it in situ} along the (11$\bar 2$0) crystal plane [see Fig. 1(b)], as confirmed by the low-energy electron-diffraction (LEED) measurement on the cleaved surface [Fig. 1(c)] and the photon-energy dependence of the band dispersion as discussed later. This indicates that the sample cleaves along the mesomeric chains [dashed lines in Fig. 1(b)], corresponding to the $k_y$-$k_z$ plane in the hexagonal BZ [Fig. 1(d)]. Sample temperature was kept at $T$ = 40 K during the ARPES measurements. The Fermi level ($E_{\rm F}$) of samples was referenced to that of a gold film evaporated onto the sample holder. Electronic band-structure calculations were carried out by means of a first-principles density-functional-theory approach by using projector augmented wave method implemented in Vienna {\it Ab initio} Simulation Package (VASP) code \cite{VASP} with generalized gradient approximation (GGA) \cite{GGA}. After the crystal structure was fully optimized, the spin-orbit coupling (SOC) was included self-consistently. It is noted that, while our GGA calculations underestimate the bulk band gap, the obtained band dispersions in the VB region agree reasonably well with the previous GW band calculation \cite{HirayamaPRL2015}, enabling the reliable discussion on the VB structure in the present study.

Figure 1(e) displays the energy distribution curve (EDC) in the wide energy region measured at the photon energy ($h\nu$) of 80 eV. One can recognize a couple of sharp core-level peaks at the binding energy ($E_{\rm B}$) of $\sim$ 40 eV, which are attributed to the Te 4$d_{3/2}$ and 4$d_{5/2}$ spin-orbit satellites, respectively. Within the $E_{\rm B}$ of 6 eV, we observed a weak intensity of the Te 5$p$ states forming the VB. No other peaks except for the Te orbitals were found in the entire energy range, confirming the clean surface. Figure 1(f) shows the EDCs at the normal emission measured with various photon energies from 52 to 98 eV. One can identify several dispersive bands. For example, a band located at $E_{\rm B}$ $\sim$ 4.7 eV at $h\nu$ = 64-68 eV disperses toward lower $E_{\rm B}$'s on increasing $h\nu$ from 68 to 86 eV, stays at $E_{\rm B}$ $\sim$ 3.5 eV at $h\nu$ = 88-92 eV and disperses back again toward higher $E_{\rm B}$'s on further increasing $h\nu$. A similar trend has been also observed for the bands located at $E_{\rm B}$ $\sim$ 3 and 2 eV at $h\nu$ $\sim$ 90 eV. The topmost VB is located at $E_{\rm B}$ $\sim$ 1 eV and it weakly disperses along this cut. Such a periodic nature of the band dispersion is more clearly visualized in the ARPES-intensity plot in Fig. 1(g), measured along the wave vector perpendicular to the sample surface ($k_x$) which corresponds to the $\Gamma$M cut in the bulk BZ [Fig. 1(d)]. One can immediately recognize that the overall experimental band dispersion shows a good agreement with the calculated bulk bands along the $\Gamma$M cut (red and blue curves), which confirms that the cleaving plane is (11$\bar 2$0). We also found no evidence for the bands whose dispersion is stationary to the $h\nu$ variation, a hallmark of the surface state, suggesting that all the observed spectral features are attributed to the bulk bands. According to the band calculation, the VB sextuplet within $\sim$ 3 eV of $E_{\rm F}$ (red curves) originates from the nonbonding $p$ states (lone pairs) and the deeper VB at the $E_{\rm B}$ $\sim$ 3-5 eV (blue curves) is derived from the bonding states, while antibonding counterpart contributes to the CB above $E_{\rm F}$ \cite{AgapitoPRL2013}.

\begin{figure*}
\includegraphics[width=7.0in]{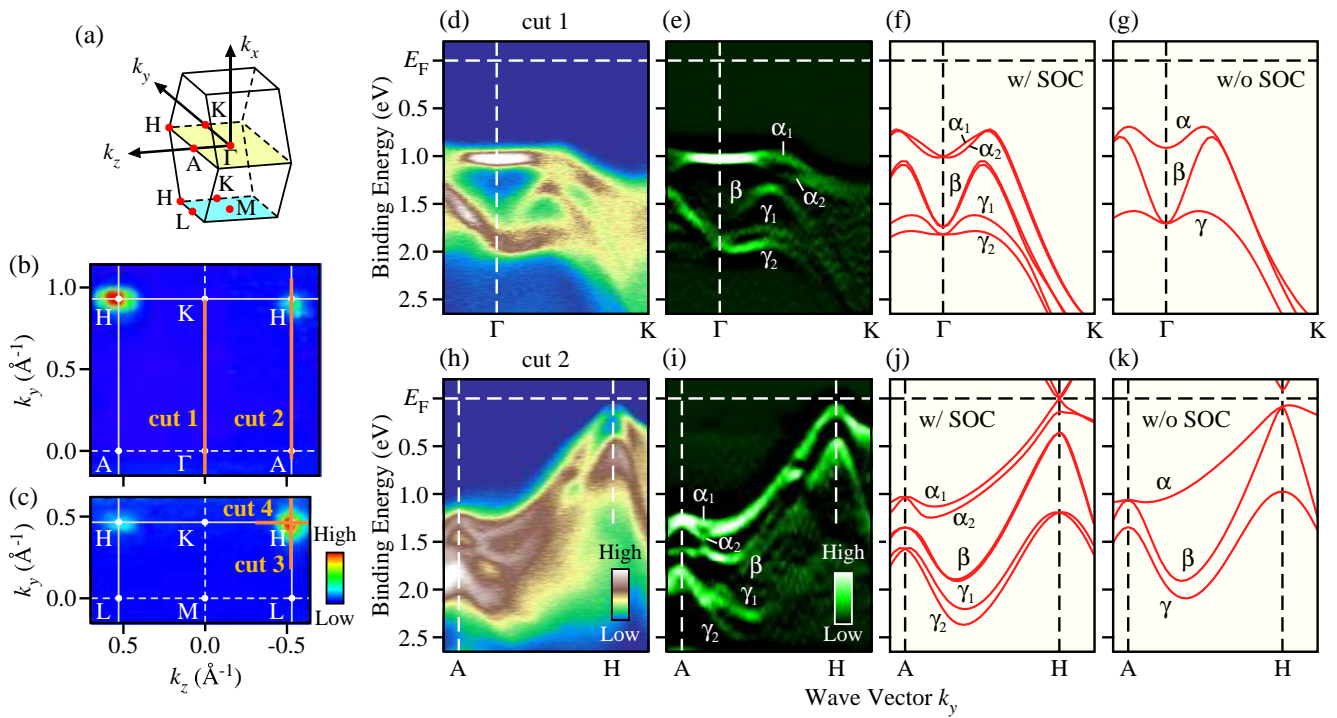}
 \vspace{-0.5cm}
 \caption{(Color online) (a) BZ of trigonal Te which is rotated with respect to Fig. 1(d). Shaded areas (yellow and blue) represent the $k_y$-$k_z$ planes where the Fermi-surface-mapping data in (b) and (c) were obtained, respectively. (b), (c) ARPES-intensity mapping at $E_{\rm F}$ on the (11$\bar 2$0) surface for $k_x$ $\sim$ 0 ($h\nu$ = 90 eV; $\Gamma$KHA plane) and $k_x$ $\sim$ $-$$\pi$ ($h\nu$ = 62 eV; MKHL plane), respectively. (d), (e) ARPES intensity and corresponding second derivative plots in the VB region, respectively, measured along cut 1 ($\Gamma$K cut) in (b). (f), (g) Calculated band dispersions for cut 1 with and without SOC, respectively. (h)-(k), Same as (d)-(g) but along cut 2 (AH cut) in (b).}
\end{figure*}

To see more clearly the lone-pair bands responsible for the various physical properties, we have chosen $h\nu$ = 90 and 62 eV photons which probe the electronic states around the $\Gamma$AHK ($k_x$ = 0) and MLHK ($k_x$ = $-$$\pi$) planes [yellow and blue shaded area in Fig. 2(a)], respectively, and mapped out the Fermi surface as a function of two-dimensional (2D) wave vector. As seen in Figs. 2(b) and 2(c), the Fermi surface consists of a small pocket at the H point, which is derived from a slight hole doping into the bulk VB top. These two independent sets of Fermi-surface mappings confirm the bulk nature of the lone-pair band crossing $E_{\rm F}$. Figures 2(d) and 2(e) show the ARPES intensity and the corresponding second-derivative intensity, respectively, measured along the $\Gamma$K high-symmetry line [cut 1 in Fig. 2(b)]. One can notice five dispersive bands (labeled $\alpha$-$\gamma$) in Fig. 2(e). The top lone-pair band at $E_{\rm B}$ $\sim$ 1 eV splits into two branches ($\alpha_1$ and $\alpha_2$) at the midway between the $\Gamma$ and K points. On approaching the K point from $\Gamma$, the middle lone-pair band ($\beta$) disperses toward lower $E_{\rm B}$ and shows a local maximum at $E_{\rm B}$ $\sim$ 1.3 eV and then disperses back toward higher $E_{\rm B}$. The bottom lone-pair bands ($\gamma_1$ and $\gamma_2$) are well separated in wide {\it k} region except for the $\Gamma$ point. These spectral features, in particular number of bands, are qualitatively reproduced by our band calculation incorporating the SOC [Fig. 2(f)], but not by that without SOC [Fig. 2(g)]. A similar behavior is also seen in the band dispersion along the AH cut shown in Figs. 2(h)-2(k). These results indicate that the strong SOC and the SIS breaking lift the band degeneracy and the observed lone-pair bands are likely spin split; this should be confirmed by spin-resolved ARPES in future. It is noted that the observed splitting is band/momentum dependent, and is maximally 0.25 eV for the $\gamma$ bands [see Fig. 2(i)], comparable to the calculated value [Fig. 2(j)]. In addition, we found no band splitting at the $\Gamma$ and A points [Figs. 2(e) and 2(i)], which is reasonable since these $k$ points are time-reversal invariant momenta (TRIM).

\begin{figure}
\includegraphics[width=3.4in]{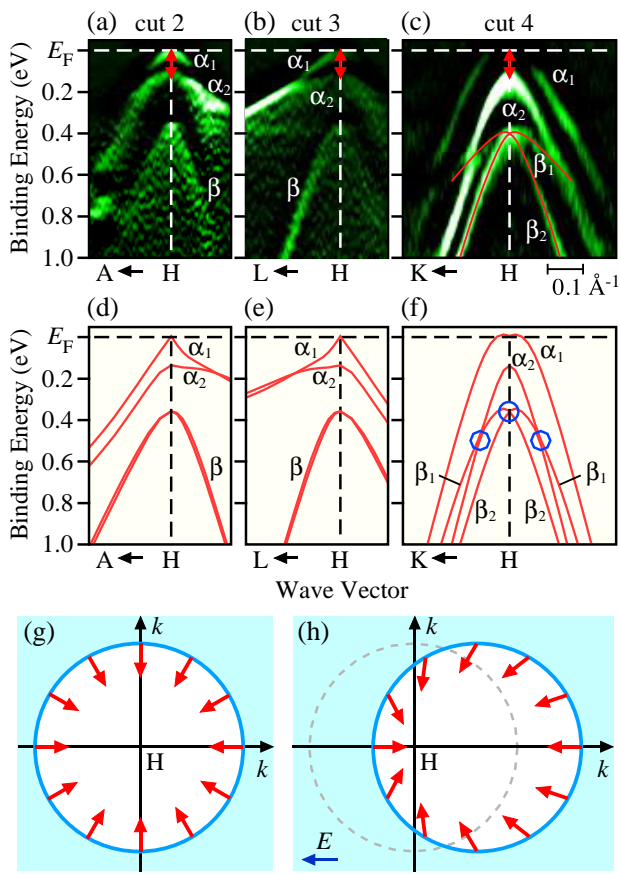}
 \vspace{-0.5cm}
 \caption{(Color online) (a)-(c) Second derivative of the near-$E_{\rm F}$ ARPES intensity crossing the H point, measured along cuts 2-4 (HA, HL, and HK cuts, respectively) in Figs. 2(b) and 2(c), compared with (d)-(f) the calculated band dispersions incorporating the SOC. Arrows in (a)-(c) represent the magnitude of the band splitting between $\alpha_1$ and $\alpha_2$ bands at the H point. Blue solid and dashed circles in (f) represent Weyl nodes with negative and positive monopole charges, respectively. (g) Schematic Fermi surface of top lone-pair band ($\alpha_1$) around the H point. Red arrows indicate calculated spin vectors. (h) Schematic Fermi surface and spin vectors of the $\alpha_1$ band under applying an electronic field along the horizontal direction.}
\end{figure}

An important consequence of the probable spin splitting of lone-pair bands is the formation of spin-polarized Fermi surface. The time-reversal symmetry does not guarantee the spin degeneracy at the H point unlike the $\Gamma$ and A points, since the H point is not a TRIM. In fact, our high-resolution ARPES data [Figs. 3(a)-3(c)] combined with the band calculations [Figs. 3(d)-3(f)] demonstrate the energy splitting of the $\alpha$ bands as large as $\sim$ 0.15 eV at the H point [see red arrows in Figs. 3(a)-3(c)], suggesting that the small hole Fermi surface arising from the $\alpha_1$ band is spin polarized. The calculated spin polarization vector around the top of the $\alpha_1$ band is schematically shown in Fig. 3(g). The spin polarization vector at $E_{\rm F}$ has dominant $x$ and $y$ components and shows radial texture (see also Figs. S1 and S2 of Supplemental Material for details of the calculated spin texture \cite{Supplementary}). Such characteristic spin texture would lead to the current-induced spin polarization due to the drift of Fermi surface and tilting of spin vectors, as seen from Fig. 3(h). Interestingly, the net spin polarization is parallel to the direction of the electric current \cite{YodaSciRep2016}, in contrast to the perpendicular polarization for the conventional Rashba spin splitting.

Another important aspect of the band dispersion around the H point is the possible presence of Weyl nodes in deeper lying lone-pair bands. As shown in Fig. 3(f), the calculated $\beta$ bands ($\beta _1$ and $\beta_2$) degenerate at the H point in contrast to the $\alpha$ bands, as indicated by blue solid circle. Such band degeneracy is also reproduced in the experiment [Fig. 3(c)]. This degeneracy is protected by the threefold screw symmetry of the helical chain and hence an important consequence of the chiral nature of crystal (both the SIS breaking and the protection by screw symmetry are required to explain this band degeneracy). More interestingly, this degeneracy point is regarded as a Weyl node with negative monopole charge at the H point \cite{HirayamaPRL2015}. The calculation also predicts another Weyl node with positive monopole charge along the HK line [blue dashed circle in Fig. 3(f)], which originates from accidental band crossing of the $\alpha_2$ and $\beta_1$ bands and resultant band degeneracy protected by the screw symmetry. In the experiment, we observed a similar band crossing between the $\alpha_2$ and $\beta_1$ bands [Fig. 3(c)].

Now we discuss implications of the present result in relation to the topological properties. In the present study, we observed the band splitting of the bulk bands associated with the SIS breaking of chiral crystal. The experimental observation of the band splitting for the {\it bulk} band in three-dimensional (3D) systems has been so far limited to only a few materials such as polar semiconductor BiTeI and noncentrosymmetric WSM TaAs families \cite{IshizakaNM2011, TaAsHongNP2015, TaAsHasanScience2015, TaAsChenNP2015}, whereas the band splitting has been widely reported in 2D systems, e.g. at surface and interface where the SIS is naturally broken. The present result unambiguously demonstrates that trigonal Te is a new member of the 3D counterpart. The observed large band splitting of up to 0.25 eV comparable to that in BiTeI and TaAs \cite{IshizakaNM2011, TaAsHongNP2015, TaAsHasanScience2015, TaAsChenNP2015} demonstrates a strong influence of the SIS breaking on the bulk electronic states. These characteristics suggest that trigonal Te (more generally, a narrow-gap semiconductor with a chiral structure and a strong SOC) is a promising platform to explore the WSM phase, since (i) the WSM phase commonly emerges around the semiconductor-to-semimetal transition point of the SIS-broken 3D system \cite{MurakamiNJP2007} and (ii) the stronger SIS breaking and the resultant large spin-splitting of bands stabilize the WSM phase over a wider range of parameter that controls the bulk-band gap \cite{MurakamiNJP2007}. To firmly establish the predicted topological phase transition, future investigation with the application of pressure or strain is of crucial importance.

Finally, we discuss the implications of the possible emergence of Weyl nodes. To the best of our knowledge, our observation of Weyl nodes in trigonal Te is the first experimental verification of the band degeneracy protected by the non-symmorphic screw symmetry. It is known that the most of symmetry-protected topological matter and gapless states have been experimentally discovered in the insulators or semimetals with the discrete symmetry like the time-reversal or the symmorphic mirror/rotational symmetry \cite{HasanReview, ZhangReview, AndoReview, LiuScience2014, BorisenkoPRL2014, NeupaneNC2014}. Recently, the non-symmorphic glide-mirror symmetry (translational plus mirror symmetry) is also recognized to realize further distinguished topological class like nodal-line semimetals and hourglass fermion phases \cite{XuPRB2015, SchoopNC2016, NeupanePRB2016, TakanePRB2016, BernevigNature2016, HongHourglass2016}. The present result experimentally establishes a new mechanism protecting the gapless states utilizing the non-symmorphic screw symmetry, and further opens a door for exploring exotic topological phases in chiral crystals.

In conclusion, we have reported ARPES results on trigonal Te and elucidated the bulk electronic states. We revealed the band splitting of the bulk bands at several locations in the BZ due to the SIS breaking of chiral crystal. We also found a good agreement of the experimental VB dispersion with the first-principles band-structure calculations, and revealed a signature of Weyl nodes protected by the non-symmorphic screw symmetry as well as the spin-polarized Fermi surface around the H point. The present result provides a pathway toward exploring new topological materials and quantum phenomena utilizing chiral crystals.

\begin{acknowledgements}
We thank D. Takane, H. Oinuma, K. Nakamura, G. Phan, S. Kanayama, Y. Nakata, N. Shimamura, N. Inami, K. Horiba, H. Kumigashira, and K. Ono for their assistance in the ARPES measurements. This work was supported by grants from JSPS KAKENHI Grant Numbers JP15H02105, JP25287079, JP15H05853, JP25107003, KEK-PF (Proposal number: 2015S2-003), and the Program for Key Interdisciplinary Research, and the Murata Science Foundation.\end{acknowledgements}

\end{document}